\documentclass[journal=langd5,manuscript=article]{achemso}

 \usepackage{float}
 \usepackage[labelfont=bf,labelsep=period]{caption}
 \usepackage[english]{babel}
 \usepackage{sectsty}
 \usepackage[utf8]{inputenc}
 \usepackage{array}
 \usepackage{geometry}
 \usepackage{setspace}
 \usepackage{xkeyval}
 \usepackage{xcolor}
 \usepackage{natbib}
 \usepackage{chngcntr}
 
 \usepackage{siunitx}

\author{Howon Choi}
\affiliation{Department of Chemical and Materials Engineering, University of Alberta, Alberta T6G 1H9, Canada}

\author{Zixiang Wei}
\affiliation{Department of Chemical and Materials Engineering, University of Alberta, Alberta T6G 1H9, Canada}

\author{Jae Bem You}
\affiliation{Department of Chemical and Materials Engineering, University of Alberta, Alberta T6G 1H9, Canada}

\author{Huaiyu Yang}
\affiliation{Department of Chemical Engineering, Loughborough University}

\author{Xuehua Zhang}
\affiliation{Department of Chemical and Materials Engineering, University of Alberta, Alberta T6G 1H9, Canada}
\alsoaffiliation{Physics of Fluids Group, Max Planck Center Twente for Complex Fluid Dynamics, JM Burgers Center for Fluid Dynamics, Mesa+, Department of Science and Technology, University of Twente, Enschede 7522 NB, The Netherlands.}
\email{xuehua.zhang@ualberta.ca}

\title {Effects of chemical and geometric micro-structures on crystallization of surface droplets during solvent exchange}

\begin{document}

\begin{abstract}



In this work, we investigate crystallization from droplets formed on micro-patterned surfaces. By solvent exchange in a micro-chamber, a ternary solution consisting of a model compound beta-alanine, water, and isopropanol, was displaced by a flow of isopropanol. In the process, oiling-out droplets formed and crystallized. Our results showed that the shape and size of the crystals on micro-patterned surfaces could be simply mediated by the flow conditions of solvent exchange. Varying flow rate, concentration, or channel height led to the formation of a thin film with micro-holes, connected network of crystals, or small diamond-shaped crystals. Rough micro-structures on the surface allowed the easy detachment of crystals from the surface. Beyond oiling-out crystallization, we demonstrated that the crystal formation from another solute dissolved in the droplets could be triggered by solvent exchange. The length of crystal fibers after the solvent exchange process was shorter at a faster flow rate. This study may provide further understanding to effectively obtain crystallization from surface droplets through the solvent exchange approach.

\end{abstract}

\section{1. Introduction}
\addcontentsline{toc}{section}{Introduction}

Crystallization from droplets is a common process in the separation and purification of compounds in solutions. The most notable example is oiling-out crystallization, a crystallization process with a secondary liquid phase appearing (liquid-liquid phase separation, LLPS) during the cooling or adding an anti-solvent\cite{kiesow_2010_solubility, lafferrre_2004_study, derdour_2010_a}. The oiling-out phenomenon was observed and reported in the crystallization of active pharmaceutical ingredients (ibuprofen\cite{codan_2012_phase}, erythromycin ethylsuccinate\cite{li_2016_process}, idebenone\cite{lu_2012_study}, ethal- propyl- and butyl- paraben\cite{yang_2010_solubility,yang_2013_nucleation}) and food components such as vanillin\cite{zhao_2012_solution, du_2016_the}, and lauric acid\cite{maeda_1997_separation}. After oiling-out, the solution becomes cloudy due to the formation of the droplets (dispersed phase) in the continuous phase but eventually separates into two layers of subphases. If the droplet phase has a high solute concentration, the oiling-out processes lead to agglomeration, lowering the quality and purity of the final crystal products\cite{deneau_2005_an,gielen_2017_agglomeration}. It is important to understand and control the phase-phase separation \cite{codan_2010_phase,sun_2018_understanding, yang_2014_sandwich} and design the oiling-out process. To control the crystal size distribution and quality of the crystals \cite{sun_2018_oilingout, yang_2014_influence, li_2016_oiling}, the crystallization processes can be designed to avoid oiling-out by lowering the initial concentration, adjusting temperature and pressure, seeding, or employing ultrasounds\cite{ogrady_2007_the, tari_2019_comparative, beck_2010_controlled, kluge_2012_emulsion}. On the other hand, some research was reported to use the oiling-out to fabricate spherical agglomeration for the convenience of formulation in downstream processing\cite{sun_2018_oilingout}. 

Our latest work demonstrated that oiling-out crystallization may provide a simple approach to prepare seed crystals from a stock solution at a very low concentration of the solute\cite{zhang_2020_oilingout}. The oiling-out crystallization of beta-alanine as a model compound occurred in a process called solvent exchange, where a good solvent (high solute solubility) was continuously replaced by a poor solvent (low solute solubility).  During the solvent exchange process, the droplets of alanine solution nucleated on the substrate and the water (the good solvent) in the droplets were continuously replaced by isopropanol (the poor solvent). With more good solvent exchanged, the droplets entered into solid-liquid-liquid phase region where the crystallization occurred\cite{zhang_2020_oilingout}. We have demonstrated that the crystal morphology and the crystallization processes on the homogeneous surfaces were largely influenced by the flow rate and concentration ratio of good and poor solvents.


It is known from the formation of ternary liquid nano-droplets that the droplet size is correlated with the initial solution concentration, flow rate, or the channel height during the solvent exchange \cite{zhang_2015_formation,yu_2015_gravitational,dyett_2018_coalescence, qian_2019_surface}. On a homogeneous surface, the nucleation and coalescence of the droplets are random, leading to the heterogeneous final size of droplets by the end of solvent exchange. However, highly ordered droplet arrays can be formed on chemically micro-patterned surfaces, as the location and spatial arrangement of the droplets follow the wettable microdomains on the surface \cite{bao_2015_highly,bao_2016_controlling, devi_2017_sessile, encarnacinescobar_2019_morphology}. For the crystallization of lipid nano-droplets in arrays, the small-angle X-ray scattering (SAXS) measurement during the cooling crystallization showed faster crystallization in comparison to bulk lipid crystallization due to the droplet morphology\cite{dyett_2018_crystallization}. We expect that the size and location of oiling-out droplets are influenced by micro-patterns on the surface, which may lead to different shapes and sizes of crystals from the solvent exchange crystallization of these droplets.

In this work, we have studied the effects of micro-patterns and micro-structures on the surface on the formation of oiling-out droplets and crystals during solvent exchange. Furthermore, we will show that solvent exchange method can induce crystallization of droplets even without the oiling-out process. Our work demonstrates that the oiling-out crystallization from droplets with solvent exchange is a potential method to screen and explore the new morphology of the crystalline product. The results in this work contribute to further understanding the crystallization from droplets in order to integrate with the application of pharmaceutical production and other future implementations.




\section{2. Experimental section}
\addcontentsline{toc}{section}{Experimental section}

\subsection{2.1 Substrate preparation}
\addcontentsline{toc}{subsection}{Substrate preparation}

The bare silicon substrates were purchased from University Wafer (South Boston, MA, US). The bare silicon wafer was used as bases for forming octadecyl trichlorosilane (OTS) coated substrate and 3-aminopropyltriethoxysilane (APTES) coated substrate. To prepare the wafer, the bare silicon wafer was cleaned by sonication in deionized water (from Milli-Q Direct) and then in ethanol for 10-20 minutes each before use. The bare silicon substrate was cleansed by piranha solution for 10-20 minutes at 75 \textdegree C by a hot plate, then sonicated by deionized water for 10-20 minutes.

Bare silicon substrates were coated by OTS by following the protocol reported in the literature\cite{zhang_2008_nanobubbles}. A similar protocol was followed for APTES coating on the substrate as well. The OTS-coated wafers were used as the bases for fabricating the patterned substrate via standard photolithography followed by plasma etching as indicated in the works of Bao et al.\cite{bao_2015_highly,lei_2018_formation}. 20 \textmu m hydrophilic domain surrounded by hydrophobic background was used as hydrophilic patterned substrates, and 5 \textmu m hydrophobic domain surrounded by hydrophilic background was used as hydrophobic patterned substrates. 1, 6-hexanediol diacrylate (HDODA) micro-lenses on the substrate were prepared by following the protocol in the literature.\cite{bao_2015_highly,lei_2018_formation}
 
 The contact angle of the water droplet in the air on APTES-coated substrate was approximately 50-60 degrees. The contact angle from hydrophilic and hydrophobic components of the patterned substrate was approximately 20-30 and 100-110 degrees respectively.

\subsection{2.2 Solution preparation}
\addcontentsline{toc}{subsection}{Solution preparation}

Beta-alanine (ACROS organics, 99 \%), ethanol (Fisher Scientific, HPLC Grade, including 90 \% ethanol, 5 \% methanol, 5 \% isopropanol), isopropanol (IPA, Fisher Scientific, 99.9 \%) were mainly used for solution preparation. The solution was stored in a sealed container overnight before use. Solution A was 3 \% beta-alanine, 45 \% water, and 52 \% isopropanol. The solutions were prepared by mixing the beta-alanine with isopropanol and water mixture followed by sonication until the solution became homogeneous. Solution B only consisted of isopropanol. All the concentration ratios are by weight percent unless indicated otherwise. 

2,5-Bis(2-(4-pyridyl)-vinylene) hydroquinone dineopentyl ether (Np-P4VB) was prepared by following the protocol reported in literature\cite{lane_2018_widegamut}. The solution of Np-P4VB was prepared similarly to the preparation of the alanine solution. The solvent ratio for solution A consisted of 15:4:0.35 (ethanol:water:mesitylene) with approximately 0.01 g of Np-P4VB per 50 mL of the solvents. A small amount of mesitylene (Sigma-Aldrich, 98 \%) was added to minimize the use of the solute as Np-P4VB was artificially created in a small quantity. The prepared solution was then stored away from the light as the solute was sensitive to light.

\subsection{2.3 Procedure of oiling-out crystallization by solvent exchange}
\addcontentsline{toc}{subsection}{Procedure of oiling-out crystallization by solvent exchange}

\begin{figure} [htp]
	\includegraphics[trim={1.5cm 4.9cm 7.7cm 2.5cm}, clip, width=0.9\columnwidth]{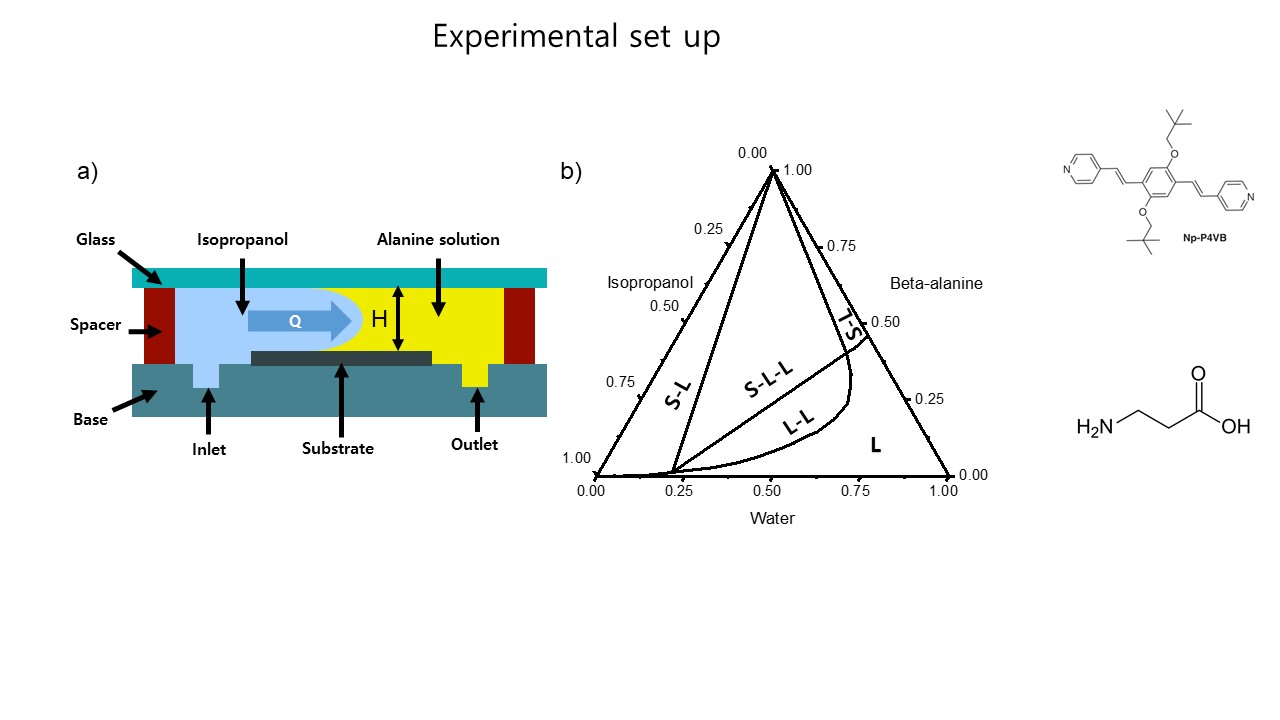}
	\caption[Experimental apparatus and ternary phase diagram]{a) Sketch of the side view of fluid chamber. The solvent exchange process is progressed as alanine solution is replaced by isopropanol. b) Ternary phase diagram of beta-alanine, isopropanol, and water at 25 \textdegree C taken from Sun et al.\cite{sun_2018_oilingout} Regions on the diagram are solid-liquid region (S-L), solid-liquid-liquid region (S-L-L), liquid-liquid region (L-L), and homogeneous liquid region (L).}
	\label{f1}
\end{figure}

The process of solvent exchange started with solution A being filled in the flow cell as shown in Figure \ref{f1} (a). Solution A was then displaced by solution B. The flow of solution B continued until all droplets crystallized and ran for 1-2 minutes extra for possible incompletion of crystallization. The syringes and flow cell were subjected to washing and sonicating by ethanol and water to avoid possible contamination or crystal leftover after each experiment. After sonication, the flow cell and syringes were dried by compressed air before assembling the flow cell for the next experiment. For the adjustment of the channel height (H), layers of double-sided tape were used to create a higher platform for the substrate to reach from 100 \textmu m to 300 \textmu m.

The ternary phase diagram of beta-alanine, water, and isopropanol is referenced in Figure \ref{f1} (b). There are five regions within the ternary phase diagram: solid-liquid region (solute lean), solid-liquid-liquid region, solid-liquid region (solute rich), liquid-liquid region, and homogeneous liquid region. The data for the phase diagram was acquired from beta-alanine experimental works from Sun et al.\cite{sun_2018_oilingout}


\subsection{2.4 Solute concentration, flow rate and channel height}
\addcontentsline{toc}{subsection}{Solute concentration, flow rate and channel height}

  The conditions for the main experiment can be viewed in Table \ref{tp2}.

\begin{table}[h!]
\centering

\begin{tabular}{ |p{1cm}|p{3cm}|p{3cm}|p{3cm}|p{3cm}|  }
 \hline
 No. & Solution A & Flow rate & channel height & Peclet number\\
 \hline
 1 & 1.5 \% alanine & 6 mL/hr & 300 \textmu m & $170$\\
 2 & 1.8 \% alanine & 6 mL/hr & 300 \textmu m & $170$\\
 3 & 3 \% alanine & 6 mL/hr & 300 \textmu m & $170$\\
 4 & 3 \% alanine & 12 mL/hr & 300 \textmu m & $330$\\
 5 & 3 \% alanine & 50 mL/hr & 300 \textmu m & $1400$\\
 6 & 3 \% alanine & 70 mL/hr & 300 \textmu m & $1900$\\
 7 & 3 \% alanine & 70 mL/hr & 200 \textmu m & $1900$\\
 8 & 3 \% alanine & 70 mL/hr & 100 \textmu m & $1900$\\
 \hline
 \multicolumn{5}{|p{13cm}|}{ Peclet number ($Pe$) is defined as $\frac{Q}{wD}$ where $Q$ is the flow rate, $w$ is the channel width, and $D$ is the diffusion constant.} \\
 \hline
\end{tabular}
\caption{List of experimental conditions.}
\label{tp2}
\end{table}

 
\subsection{2.5 Data collection}
\addcontentsline{toc}{subsection}{Data collection}

The videos and images for the progress of oiling-out crystallization in the flow cell during the solvent exchange were accomplished by NIKON Eclipse Ni microscope with X-Cite Series 120 Q mercury light source. The NIKON software has an auto-whiting function which corrected the color saturation and an auto-exposure correction that adjusts the brightness of the videos and images. Both of these features were utilized for our experiments. We selected video frames with a fixed interval based on the rate in droplet and crystal growth.
 
The initial time $t_{0}$ in our experiments is defined as the time at the start of droplet formation on the substrate. There are few droplets shown in the bulk flow during the solvent exchange before $t_{0}$ but the widespread droplet formation on the substrate occurred separately. We based our time stamps only on the observation of the substrates through an optical microscope.
 
\section{3. Results and Discussion}

Effects of chemical patterns on the oiling-out crystallization are coupled with the influences from other conditions including hydrophobicity/hydrophilicity of the micro-patterns, flow rate, channel geometry, and solute concentration. We will show effects from chemical micro-patterns on smooth surfaces. Finally, we will demonstrate that solvent exchange can be used to induce crystallization of a solute other than alanine.

\subsection{3.1 Oiling-out crystallization on surfaces with chemical micro-patterns}
\subsubsection{a. Hydrophobic or hydrophilic micro-patterns }

\begin{figure} [htp]
	\includegraphics[trim={0.5cm 0.5cm 1.5cm 0cm}, clip, width=1\columnwidth]{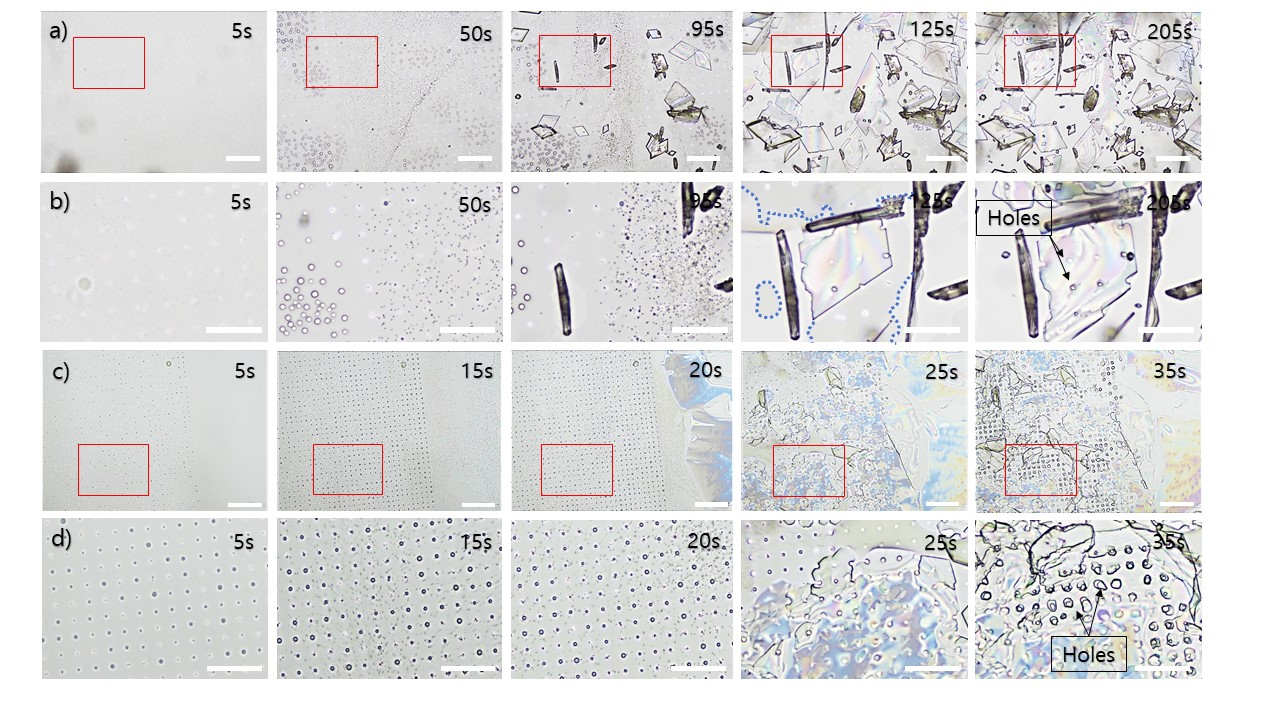}
	\caption[Oiling-out crystallization on the hydrophobic patterned substrate]{Oiling-out crystallization on the hydrophobic patterned substrate. a) At 12 mL/hr flow rate, b) the zoomed in images of the red highlighted square in a). The blue dotted line indicates the boundary of thin liquid film. c) At 70 mL/hr flow rate, d) the zoomed in images of the red highlighted square in c). Channel height: 300 \textmu m. Length of the scale bar: a) and c) 100 \textmu m and  b) and d) 50 \textmu m.
}	
	\label{f6a}
\end{figure}

During the solvent exchange at a flow rate of 12 mL/hr, a thin layer of liquid formed in the hydrophilic surrounding area while some small droplets formed on the hydrophobic micro-patterns as shown in Figure \ref{f6a} (a)(b).
The crystals evolved to a needle-like shape or a thin film similar to crystallization on homogeneous hydrophilic surfaces by solvent exchange\cite{zhang_2020_oilingout}. In the cases where the crystallization of a thin film occurred, small holes formed at the locations where the hydrophobic domains lied.

 At a faster flow rate of 70 mL/hr, droplets formed a regular array, and the holes became more apparent where the thin crystal film propagated along the surface (Figure \ref{f6a} (c)(d)). The position of holes copied the arrangement of patterns. Peng et al.\cite{peng_2017_morphological, peng_2017_simple} showed that the formation of droplets could be controlled by the patterned substrate due to preferential nucleation of oil droplets on hydrophobic areas. The alanine-rich sub-phase from oiling-out appeared to spread on the hydrophilic areas but not on the hydrophobic domains; therefore, the local rupture of the film led to the hole formation on the crystal film. This result suggested that we may make different hole arrangements on the crystal film by changing the pattern position. In particular, the size of crystal film is large at a faster flow rate, and the hole array is more regular. Such holes may allow gases, chemicals, or particles to permeate through the film. 

\begin{figure} [htp]
	\includegraphics[trim={1.2cm 6.7cm 1.0cm 3cm}, clip, width=1\columnwidth]{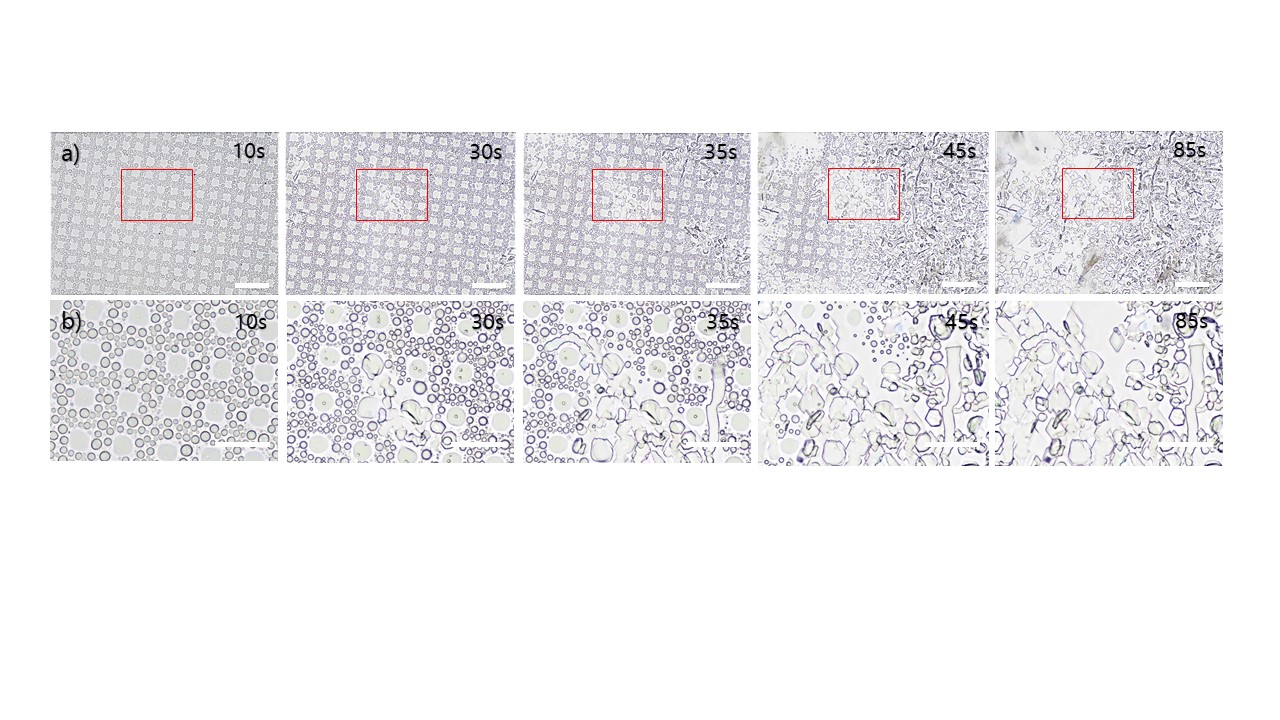}
	\caption[Oiling-out crystallization on the hydrophilic patterned substrates.]{Oiling-out crystallization on the patterned substrates. a) on hydrophilic micro-patterns b) the zoomed in images of the red highlighted square in a). Length of the scale bar: a) 100 \textmu m and b) 50 \textmu m. The condition for solvent exchange is listed as No. 6 in Table \ref{tp2}. 
}	
	\label{f6b}
\end{figure}

The oiling-out crystallization behaviour on the substrate with hydrophilic pattern and hydrophobic background is shown in Figure \ref{f6b} (a)(b). 
The droplets in hydrophilic domains are large with uniform droplet size, possibly due to the constraints from the patterns on the surface. Smaller droplets nucleated on the hydrophobic surrounding area with less uniform size distribution. The crystallization process propagated in the path of the droplet coalescence. The crystals are irregularly shaped, and the network of crystals is related to the hydrophilic patterns. The time from the initial crystallization to completion of crystallization in the field of view is around 50 s, almost twice the time on hydrophobic patterns in Figure \ref{f6a} (a)(b). A longer time for the crystal growth may be due to more droplets on the surface going through phase separation, requiring more solvent supplied from the flow.

\begin{figure} [htp]
	\includegraphics[trim={4.6cm 7.5cm 4.3cm 2.6cm}, clip, width=1\columnwidth]{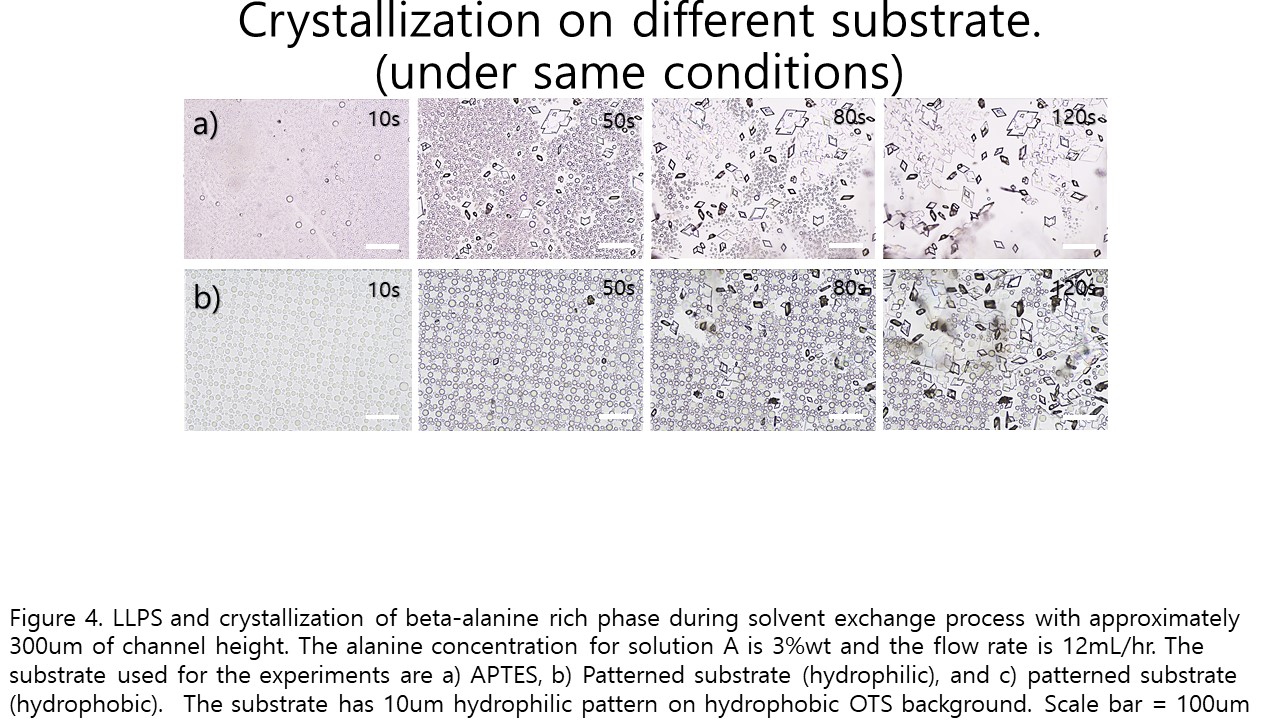}
	\caption[Oiling-out crystallization of droplets on homogeneous vs patterned substrates]{Oiling-out crystallization of droplets on a) homogeneous surface, b) hydrophilic patterns and hydrophobic surrounding area. Length of the scale bar: 100 \textmu m. The condition for solvent exchange is listed as No. 3 in Table \ref{tp2}.
}	
	\label{f6c}
\end{figure}

At a slower flow rate of 6 mL/hr in Figure \ref{f6c} (b), the hydrophilic patterns on the substrate had an even clearer effect on the droplet formation and crystallization. The position and shape of the droplets follow the pattern of the hydrophilic domains on the surface which in turn led to numerous and smaller crystals. In contrast, on a homogeneous APTES-coated substrate, the size of the crystals was much larger on average. Further studies on pattern sizes or different chemical patterns may allow optimization of solvent exchange for a specific crystal morphology and pattern.


\begin{figure} [htp]
	\includegraphics[trim={1.0cm 10cm 1.0cm 4.3cm}, clip, width=1\columnwidth]{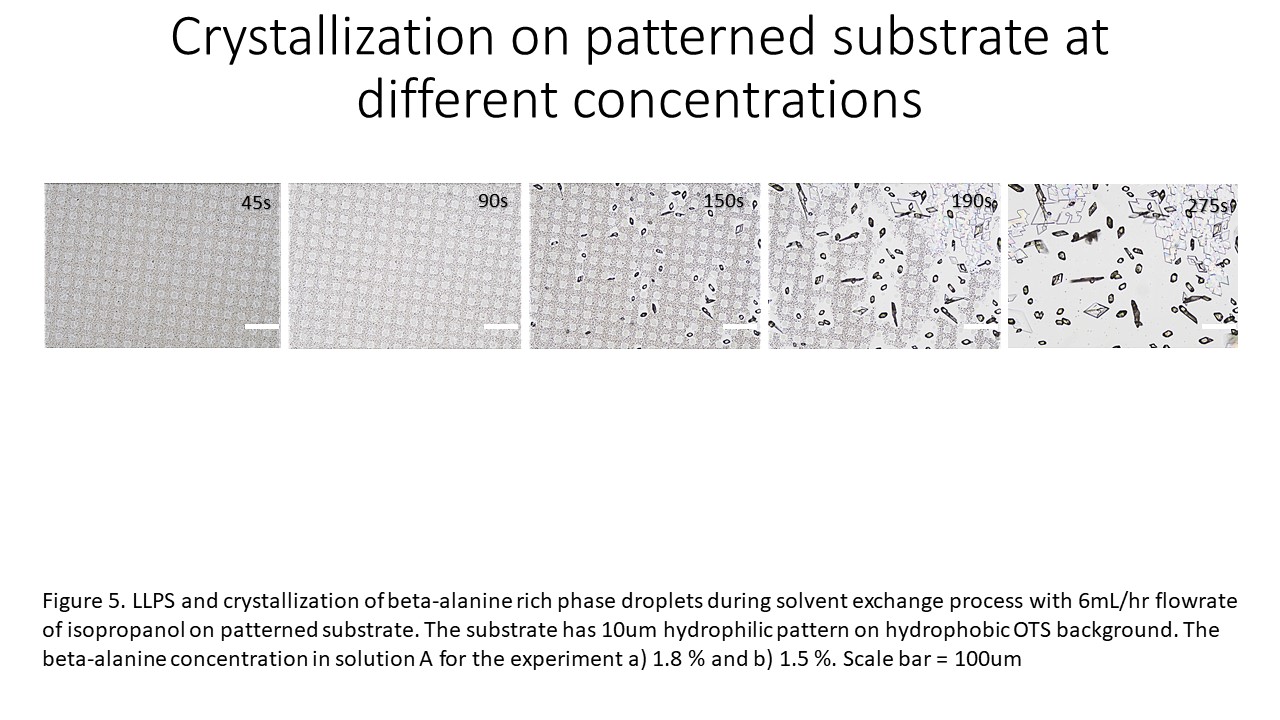}
	\caption[Oiling-out crystallization with a low alanine concentration]{Oiling-out crystallization with the low alanine concentration. The condition for the solvent exchange is No. 2 in Table \ref{tp2}. Length of the scale bar: 100 \textmu m.
}	
	\label{f6}
\end{figure}

The initial concentration of the solute also has an important effect on the droplet formation in the solvent exchange system. At an even lower concentration at 1.8 \% alanine in solution A, Figure \ref{f6} shows the decrease in crystals in comparison to the concentration of 3 \% in Figure \ref{f5} (a). However, the features in the droplet formation and the crystallization are similar to each other. The main differences are the smaller surface droplet coverage and the crystallization at a lower concentration of alanine, suggesting that there is a threshold of initial concentration required for oiling-out crystallization by solvent exchange.

\subsubsection{b. Flow rate of the poor solvent}

\begin{figure} [htp]
	\includegraphics[trim={1.7cm 3cm 1.5cm 2.6cm}, clip, width=1\columnwidth]{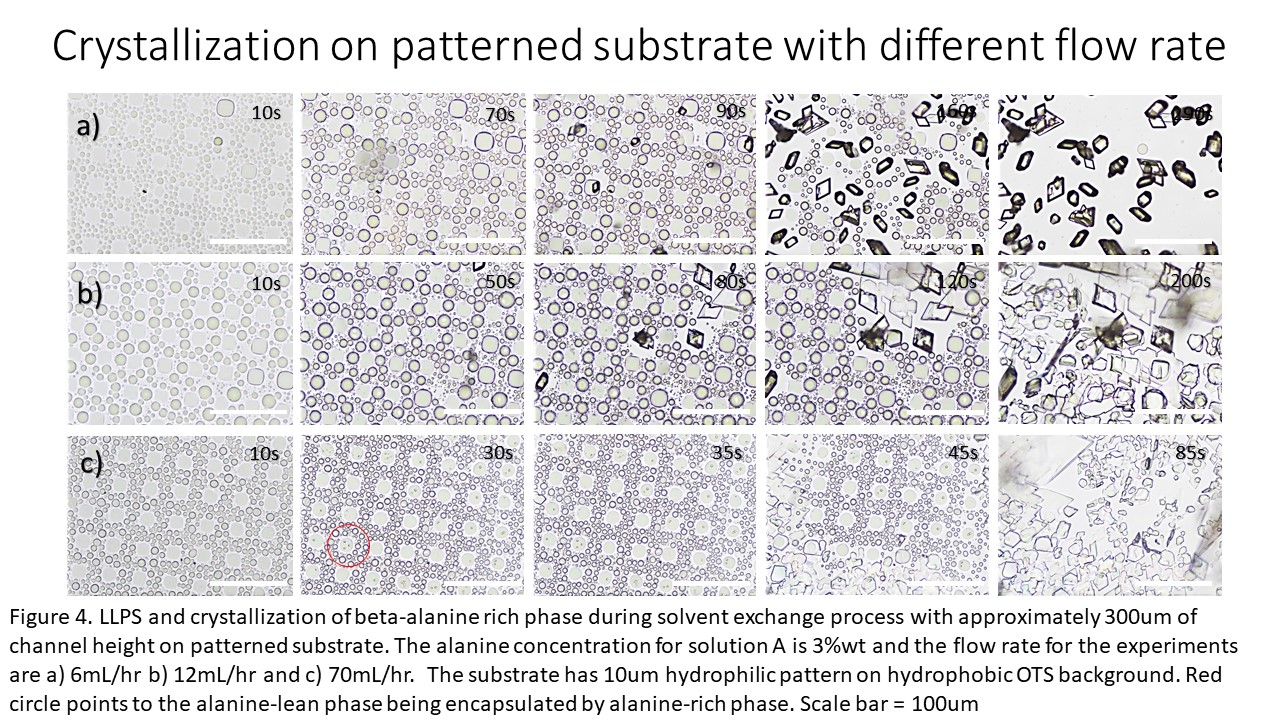}
	\caption[Oiling-out crystallization at different flow rate]{Oiling-out crystallization at the flow rate of a) 6 mL/hr, b) 12 mL/hr, and c) 70 mL/hr. The red circle indicates a case where a alanine-lean droplets are encapsulated by the alanine-rich phase. Length of the scale bar: 100 \textmu m.
}	
	\label{f5}
\end{figure}

In Figure \ref{f5}, the channel height is kept constant while the flow rate varies. The droplets at lower flow rates lasted longer before crystallization and were more likely to dissolve instead of coming to full crystallization. The completion of droplet depletion took around 290 seconds for 6 mL/hr, 200 seconds for 12 mL/hr, and 85 seconds for 70 mL/hr. All three flow rates had a single droplet that covered the hydrophilic area, suggesting that the patterns effectively controlled the droplet size regardless of the flow rate. 

At the slowest flow rate of 6 mL/hr, crystals on the surface had more polygonal shapes. Some crystals may agglomerate to form clusters of crystals. For 12 mL/hr, more flat crystals grew in diamond shape while still having few polygonal shapes. At the fastest flow rate of 70 mL/hr, the droplets crystallized in a network that spreads in a radiating fashion. These crystals were more irregular in shape, and the crystallization followed the path of the droplet coalescence. Different from crystallization on homogeneous surfaces, a faster flow rate led to large quantities of crystals of the droplet shape.\cite{zhang_2020_oilingout}

\begin{figure} [htp]
	\includegraphics[trim={1.7cm 7.3cm 1.5cm 2.6cm}, clip, width=1\columnwidth]{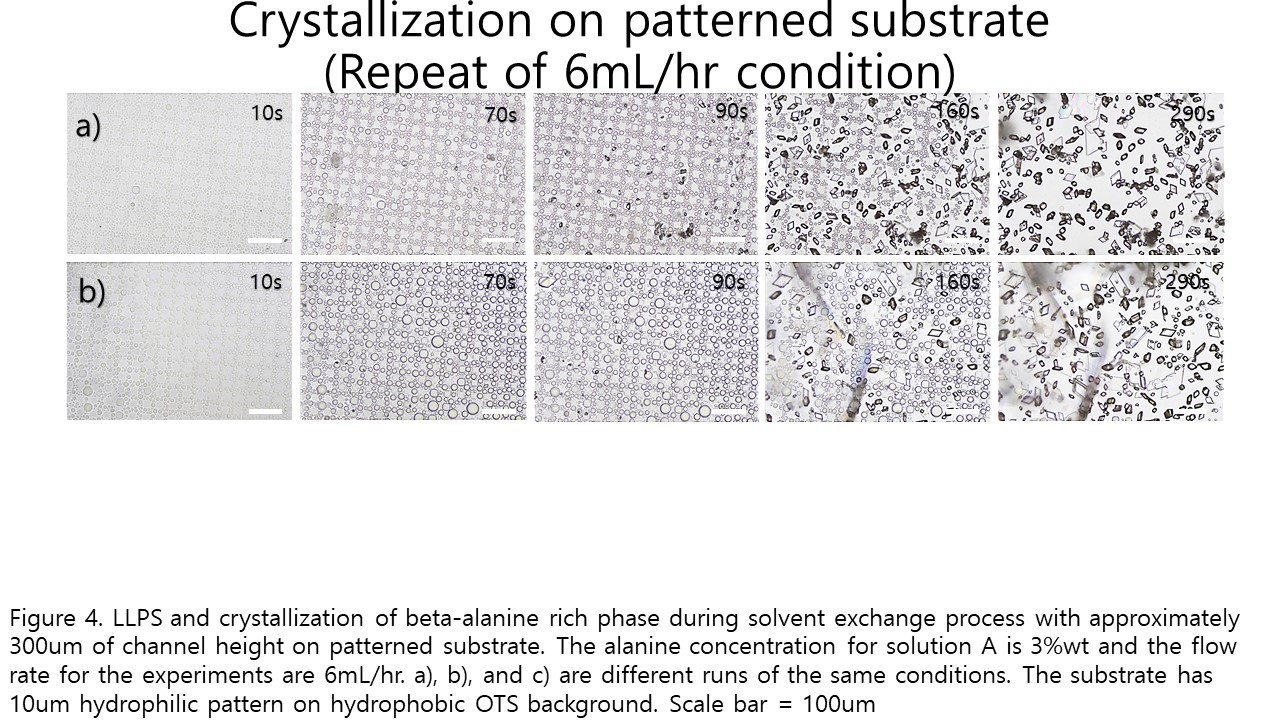}
	\caption[Reproducibility of oiling-out crystallization]{Oiling-out crystallization at the flow rate of 6 mL/hr. a) and b) are the repeats of the experiments at condition No. 3 in Table \ref{tp2}. Length of the scale bar: 100 \textmu m. 
}	
	\label{f5a}
\end{figure}

The general features in crystallization were reproducible under the same experimental conditions. As shown in Figure \ref{f5a}, some parts of the substrate or different experimental repeats may exhibit more flat crystal growth with diamond shapes. The crystals formed on patterned surfaces are much less polydispersed compared to those on a homogeneous surface. The number of crystals per $mm^{2}$ detected by the software came out to be 391 and 401 for Figure \ref{f5a} (a) and (b), respectively. Again, this indicated that general features of oiling-out crystallization were reproducible.  

Faster crystallization at a faster flow rate of solvent exchange was explained by the transport of good solvent out from the droplets\cite{zhang_2015_formation, zhang_2020_oilingout}. In brief, water diffuses out from the droplets sooner at higher flow rates, leading to the supersaturation of alanine in the droplets in a shorter time. 


The flow rate can also influence the crystallization by varying the number of the crystal seeds that land on the substrates, which is significantly greater at lower flow rates as shown in Figure \ref{f5} (a). More seeds led to smaller crystals at lower flow rates compared to larger connected crystals at higher flow rates.  Two reasons may be at play for more crystal seeds at lower flow rates. One may be due to a later crystal landing time. The droplets in the flow experienced counter diffusion, which triggers the crystallization as observed for spherical alanine crystal formation from droplets in work by Sun et al.\cite{sun_2018_understanding} The longer retention time for the ternary droplets in the flow may form more crystals in the flow before reaching a certain location on the substrate. Secondly, the slower flow rate allows more crystals to settle onto the surface in competition with advection by the bulk flow. Therefore, more crystal seeds can land on the substrate during the solvent exchange process.

\subsubsection{c. Height of the channel for solvent exchange}

\begin{figure} [htp]
     \centering
         \includegraphics[trim={0.5cm 0cm 2cm 1cm}, clip, width=1\columnwidth]{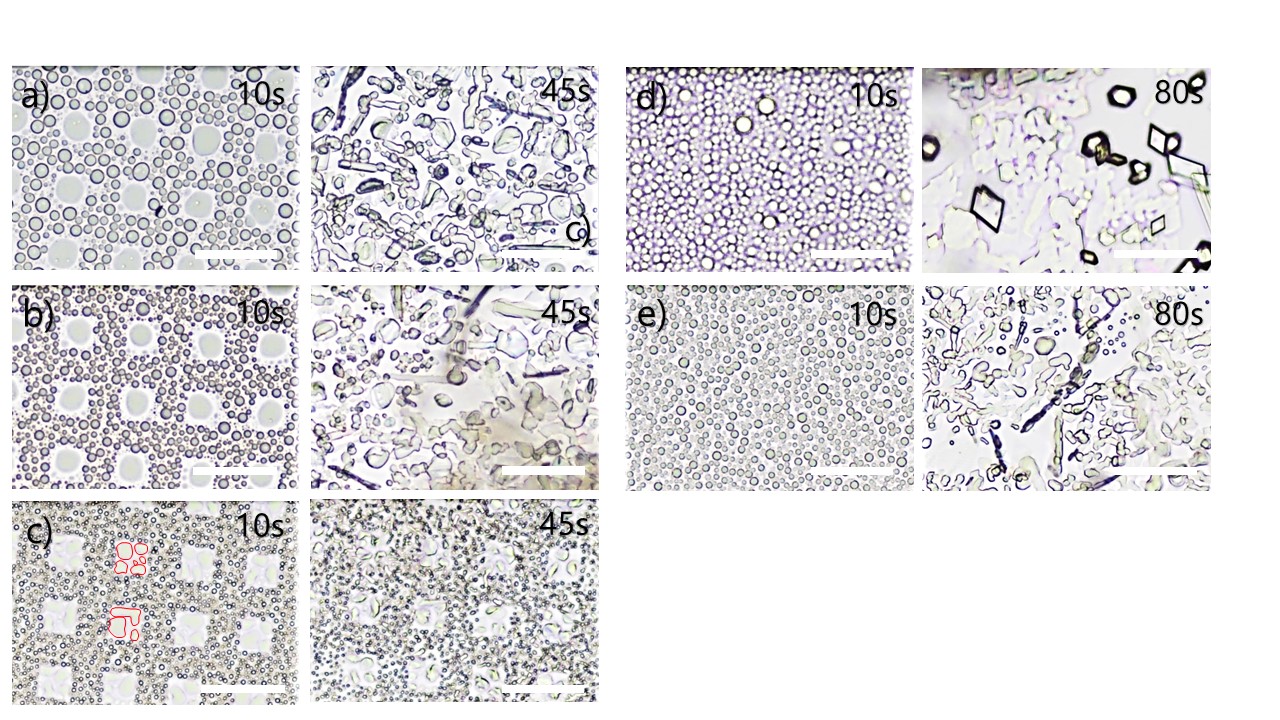}

	\caption[Decreasing droplet size with decrease in channel heights]{Decreasing droplet size with decrease in channel heights. Droplets and crystals on the patterned substrate at a) 300 \textmu m, b) 200 \textmu m, and c) 100 \textmu m and on the homogeneous substrate at channel height of d) 300 \textmu m and e) 200 \textmu m. The red outline indicates the droplet outline in the hydrophilic domain. Length of the scale bar: 50 \textmu m.
}	
	\label{fpd}
\end{figure}

 The crystallization from the solvent exchange performed in the channels with three different heights is shown in Figure \ref{fpd}. The substrates with hydrophilic micro-patterns surrounded by hydrophobic areas were used in the experiment (Figure \ref{fpd} (a)-(c)). At a given flow rate of 70 mL/hr, the droplet size decreased with a decrease in the channel height. There was only one droplet per micro-domain at the channel height of 200 \textmu m or 300 \textmu m, but multiple droplets at the channel height of 100 \textmu m.
 
  As a comparison, effects from channel height were also examined on homogeneous APTES-coated substrate (Figure \ref{fpd} (d)(e)). Crystallization followed the droplet formation with the channel height of 200 \textmu m and 300 \textmu m. When channel height was down to 100 \textmu m, no droplet formation was observed. The droplets decreased in size with the decrease in channel heights. The dependence of droplet sizes on the channel height is consistent with that on the micro-patterned substrate. 


The previous work showed that by solvent exchange, the final volume of the droplet $V_f$ increased with the channel height $h$\cite{zhang_2015_formation,yu_2015_gravitational}. At the same flow rate (in volume per unit time), $V_{f} \sim h^{3}$ because $h$ determines how long it takes for the solution A and B to mix uniformly across the channel. The higher the channel, the longer the time for the droplets to grow \cite{zhang_2015_formation,yu_2015_gravitational}.  

Above certain channel height, the effect of gravity on the mixing between two solutions may also become important. Archimedes number $Ar$ is the dimensionless number to describe the gravity effect, arising from the density difference between solution A and the displacing solution \cite{yu_2015_gravitational}.  

\begin{equation}
Ar = \frac{g (h)^3}{\nu^{2}}\frac{\Delta \rho}{\rho}
\label{eq1}
\end{equation}

The density of the solvents in solution A is 0.885 $g/mL$, while solution B is 0.786 $g/mL$. The density difference $\Delta$ $\rho$ is $\sim$ 0.1 $g/mL$, the gravitational acceleration $g$ is 9.8 $m/s^2$, and viscosity $\nu$ is $2.8 \times 10^{-6} m^{2}/s$. For our channel height $h$ of 100, 200, and 300 \textmu m, the $Ar$ number is 0.16, 1.3, and 4.3, respectively. For such a large Archimedes number ($Ar > 1$) at the channel height $h$ of 300 \textmu m, the center of the parabolic mixing front in the laminar flow is shifted towards the bottom surface under the gravity. Such shift would lead to a longer growth time for droplets and crystals on the top surface than on the bottom surfaces of a horizontal channel. However, the gravity effect could be eliminated by placing the channel vertically \cite{yu_2015_gravitational}. 

Another effect that may arise in a higher channel is enhanced mixing from convection \cite{zhang_2015_formation,yu_2015_gravitational}. 
To estimate when the convection rolls set in due to the density difference, we calculate Rayleigh number $Ra$.
\begin{equation}
Ra=\frac{\Delta \rho g (h/2)^3}{\mu D}
\label{eq2}
\end{equation}
where $\mu$ is the dynamic viscosity of solution A and the mass diffusion coefficient $D$ is approximately $10^{-5} cm^2/s$. For the channel height $h$ of 100, 200, and 300 \textmu m, $Ra$ number is 71, 570, 1900, respectively. The convection rolls may occur for 300 \textmu m channel heights where the Rayleigh number is slightly above the critical Rayleigh number 1708\cite{faber_1995_fluid}. These convection rolls enhanced mixing conditions and led to larger droplets along with the rolls than ones away from the rolls.

\subsection{3.2 Oiling-out crystallization on micro-structured surfaces}

\begin{figure} [htp]
	\includegraphics[trim={1.7cm 4.3cm 1.5cm 2.6cm}, clip, width=1\columnwidth]{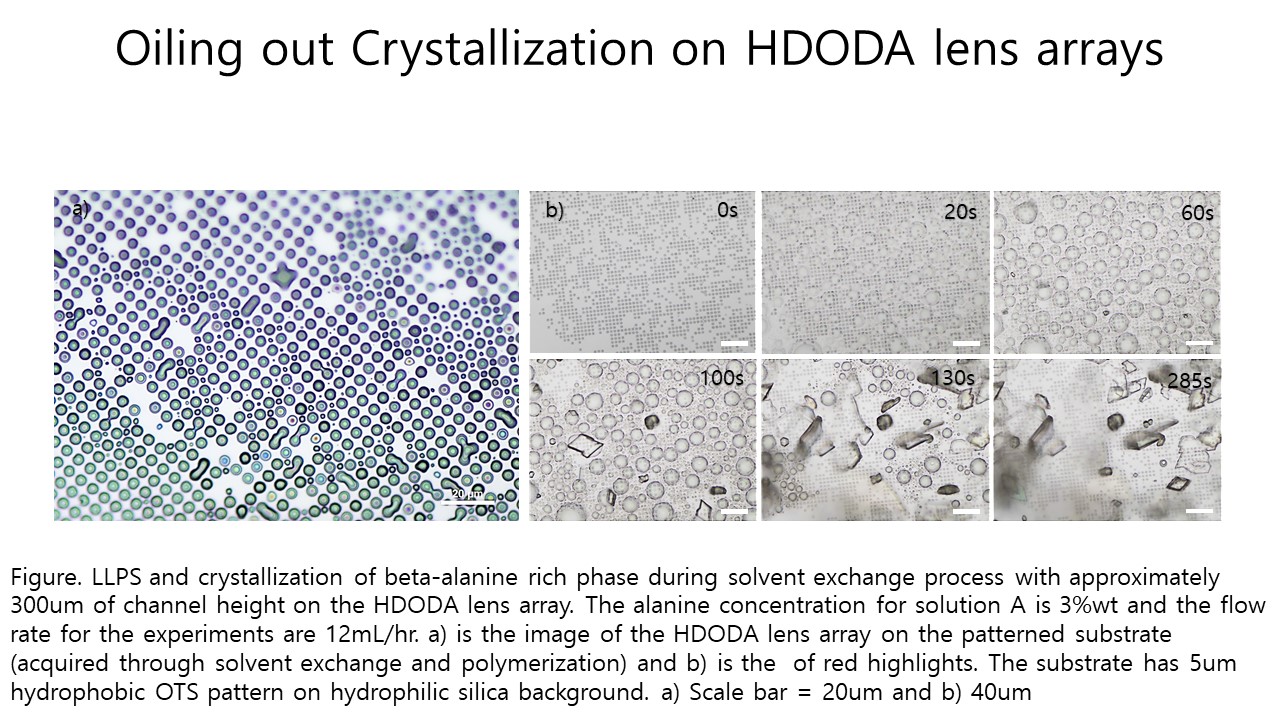}
	\caption[Oiling-out crystallization on the lens array]{Oiling-out crystallization on the lens array. a) an image of the lens array on the substrate b) the snapshot of the solvent exchange process. Length of the scale bar: a) 20 \textmu m and b) 40 \textmu m. Channel height: 300 \textmu m.
}	
	\label{f2a}
\end{figure}

 A substrate with the polymeric micro-lens array was used in our experiments as shown in Figure \ref{f2a} (a). These micro-lenses could be easily prepared by polymerization of surface droplets, therefore chosen as representative surface micro-structures to reveal the influence of physical structures on oiling-out crystallization. Figure \ref{f2a} b) from $t_0$ to $t_0$ + 60 s showed progression of droplet formation where $t_0$ is defined as a timestamp before observable oiling-out. With micro-lenses on the surface, the number density of droplets is much lower than that on a homogeneous surface, possibly due to fewer nucleation sites on the micro-lenses (Figure \ref{fpd} (d)). The droplets ranged from 5.6 \textmu m to 14.2 \textmu m, much larger than droplets on a homogeneous surface. The effect of the geometric structures on oiling-out droplets appears to be similar to that on the oil droplet formation by solvent exchange\cite{dyett_2018_coalescence,peng_2016_how}.

More importantly, the micro-lenses have a significant effect on the crystallization process shown in Figure \ref{f2a} (b). The crystals were easily detached from the substrate, and the crystal shapes were more irregular in comparison to those on the smooth substrates.\cite{zhang_2020_oilingout} The easier detachment of crystals may be attributed to the weak adhesion between the crystal plates and the spherical-cap shaped lenses on the substrate. In addition, the lenses caused the crystals to grow in a slightly elevated manner, which also favours crystal detachment. The self-detachment of crystals during the solvent exchange makes it possible for the collection of the crystals from the flow at the exit of the chamber. These collected crystals may be useful as seeds to trigger crystallization in the bulk crystallization\cite{zhang_2020_oilingout}.

\subsection{3.3 Fiber crystal formation from droplets without oiling-out}

\begin{figure} [htp]
	\includegraphics[trim={0cm 0cm 0cm 0cm}, clip, width=1\columnwidth]{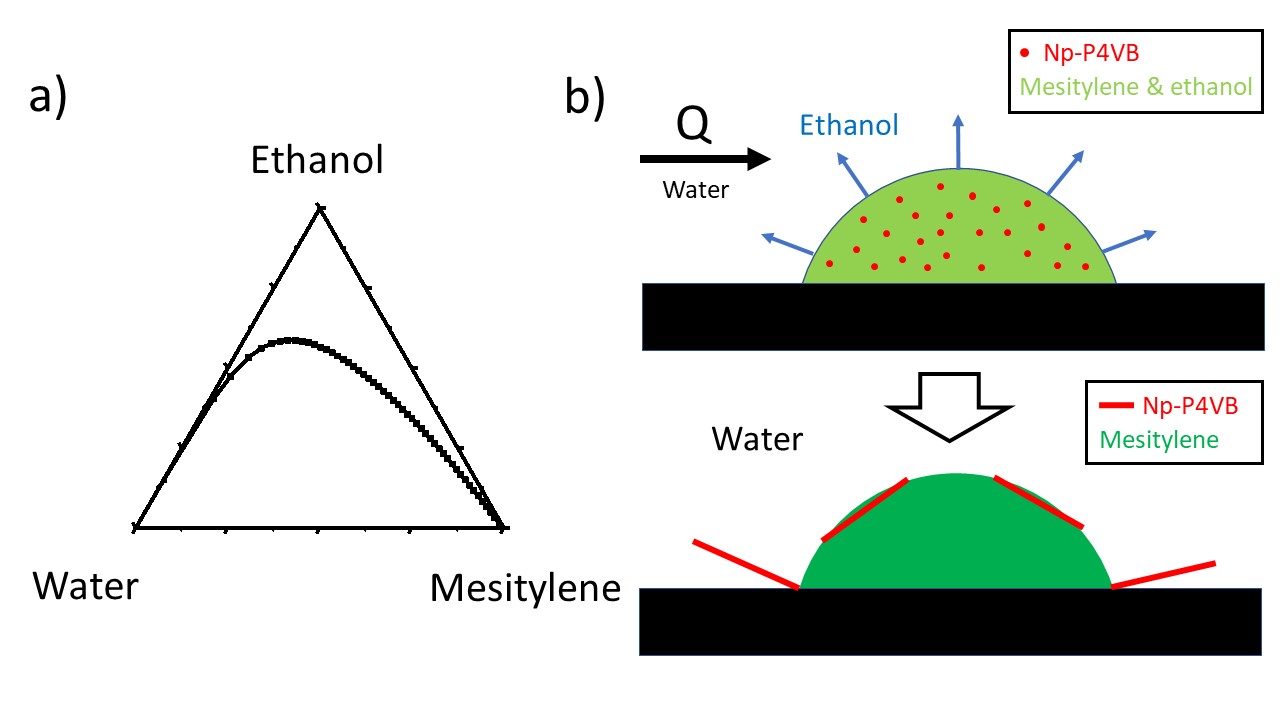}
	\caption[Ternary phase of mesitylene, ethanol, and water and a diagram of fiber formation in surface droplet]{a) Ternary phase diagram of mesitylene, water, and ethanol from IUPAC-NIST solubility database\cite{skrzecz_1997_solubility} b) Sketch of Np-P4VB fiber formation during solvent exchange.
}	
	\label{fdrawfib}
\end{figure}

Finally, we will show that the crystal formation by solvent exchange can be applied beyond the oiling-out systems. Np-P4VB was used as the solute for crystallization from droplets consisting of mesitylene, ethanol, and water. The solubility of Np-P4VB is the highest in ethanol and is the lowest in water. During the solvent exchange, solution A contained Np-P4VB, mesitylene, ethanol, and water, and solution B was water as presented in Figure \ref{fdrawfib} (a). Initially, surface droplets were mainly mesitylene with some ethanol and Np-P4VB. As the solvent exchange continued, the droplets became saturated with Np-P4VB as ethanol diffused out into the flow of water and crystallization of Np-P4VB was induced inside the droplets (Figure \ref{fdrawfib} (b)). The crystallization continued until the solute in the droplets was depleted.

\begin{figure} [htp]
\centering
	\includegraphics[trim={1.1cm 0.5cm 5cm 3cm}, clip, width=1\columnwidth]{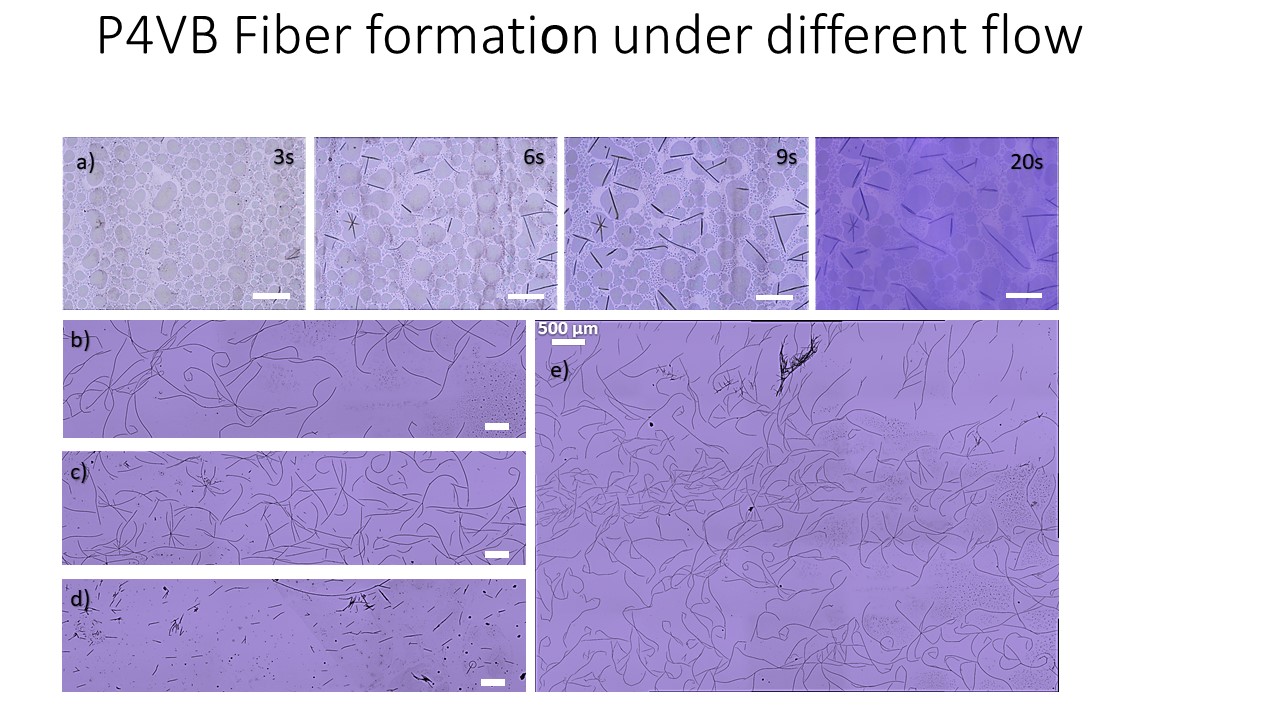}
	\caption[Fiber formation in droplet on OTS substrate through solvent exchange]{Fiber formation in droplet at flow rate of a) 100 mL/hr on OTS substrate through solvent exchange. $t_0$ is defined as the time when the droplet formation started. Crystal fiber on the substrate after solvent exchange at flow rate of b) 12 mL/hr c) 75 mL/hr d) 100 mL/hr. Length of the scale bar: a) 100 \textmu m b-d) 250 \textmu m. e) Stitched images of crystal fiber formed for b). Total of 25 images were stitched together with 10 \% overlapping to manually stitch the images.
}	
	\label{ffg}
\end{figure}


The crystal of Np-P4VB formed fibers during the solvent exchange. At a flow rate of 100 mL/hr, the fibers formed inside the droplets and extended with the solvent exchange process as shown in Figure \ref{ffg} (a). Different from the oiling-out crystallization of alanine, the crystallization initiated from the solute-rich droplets instead of the landed seed crystals, evident from the location of the crystal fibers. The onset of the crystals suggests that supersaturation may be from the reduced ethanol concentration in the droplets during the solvent exchange.

 The flow rate affected the crystal fiber formation, and longer fibers were formed at lower flow rates (Figure \ref{ffg} (b)-(d)). For a faster flow rate of 100 mL/hr, there was a significant decrease in fiber length (Figure \ref{ffg} (d)). To capture the bigger picture of the fiber formed on the substrate, the images of the fiber after solvent exchange at 12 mL/hr flow rate have been stitched as a larger image (Figure \ref{ffg} (e)). The results demonstrate that solvent exchange can be applied to control crystallization from droplets containing the solute, not limited to oiling-out systems.
 
\section{Conclusion}

We show that surface micro-patterns can be an effective way to control oiling-out crystallization during solvent exchange. A thin film of crystals can be created with holes on the substrates with hydrophobic micro-patterns. More uniform crystals form on the surfaces with hydrophilic micro-domains, as compared to a homogeneous surfaces. The channel height and flow rate of solvent exchange can be varied to control the oiling-out droplets and crystallization. By using representative micro-structures of polymeric micro-lenses on a substrate, the crystals detached easily from the surface, which may be used for collecting seed crystals to trigger crystallization in the bulk crystallization. Beyond oiling-out systems, solvent exchange can be used to induce crystallization by forming droplets containing the solute that does not exhibit oiling-out behaviour.

The results in the study show great potential in using solvent exchange for crystallization from droplets on the solid surface. Further studies in other chemicals will be valuable for using solvent exchange as a way to control, separate, and purify the crystal product for pharmaceutical and many other applications.

\section{Acknowledgements}

The project is supported by the Natural Science and Engineering Research Council of Canada (NSERC) and Future Energy Systems (Canada First Research Excellence Fund). This research was undertaken, in part, thanks to funding from the Canada Research Chairs program. We also thank Al Meldrum and Sergey Vagin for generously supplying us with Np-P4VB for our experiments.
\cite{codan_2010_phase,codan_2012_phase}

\bibliography{combination.bib}
\newpage






\newpage

\section{For Table of Contents Only}

\begin{figure} [htp]
	\includegraphics[trim={1.7cm 5cm 3cm 5cm}, clip, width=8.25 cm]{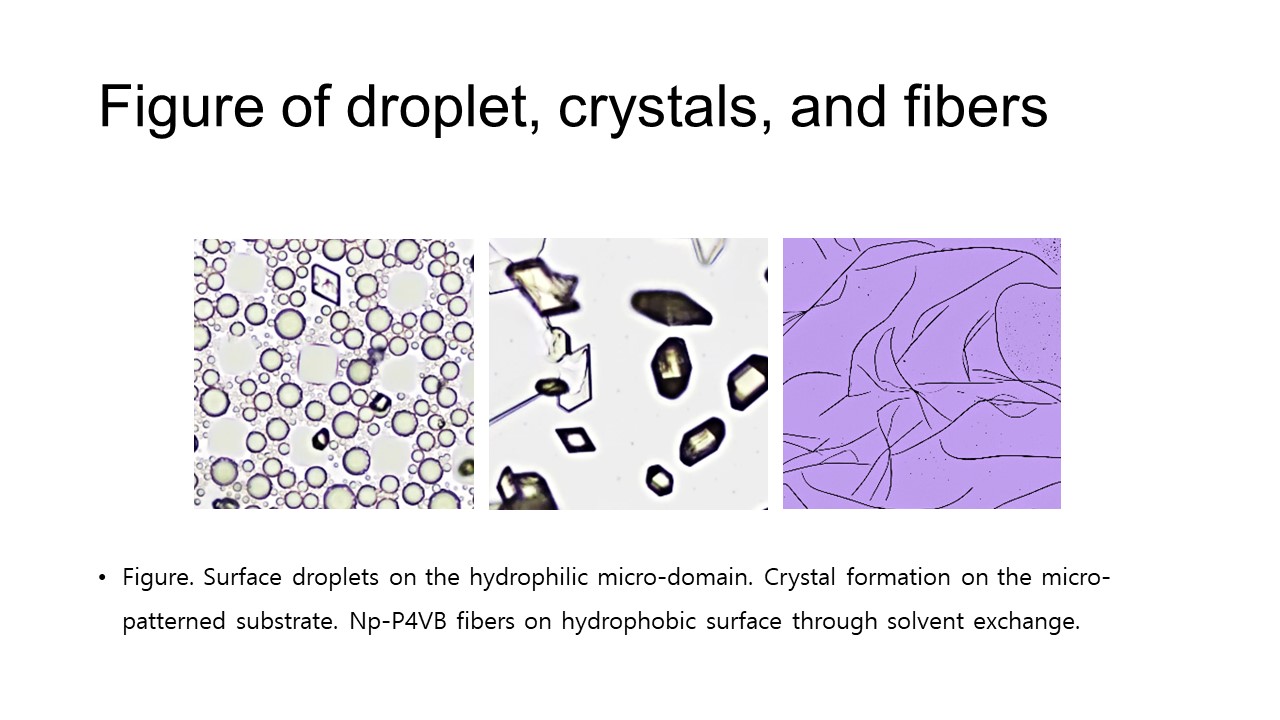}
	\label{f toc}
\end{figure}

\end{document}